# Akciğer Nodüllerinin Evrişimsel Sinir Ağları Kullanılarak Sınıflandırılmasında Girdi Boyutunun Etkisi

# Effect of Input Size on the Classification of Lung Nodules Using Convolutional Neural Networks

Görkem Polat, Yeşim Serinağaoğlu Doğrusöz, Uğur Halıcı
Elektrik-Elektronik Mühendisliği, Orta Doğu Teknik Üniversitesi
Ankara, Türkiye
gorkem.polat@metu.edu.tr, yserin@metu.edu.tr, halici@metu.edu.tr

*Özetçe—* Son yapılan çalışmalar, her yıl yapılan düşük dozlu bilgisayarlı tomografi (BT) taramalarının, geleneksel göğüs radyolojisine göre akciğer kanserinin erken tespitinde %20 daha iyi sonuç verdiğini göstermiştir. Bu sebeple akciğerin BT ile incelenmesi tüm dünyada yaygınlaşmaktadır. Fakat bu görüntülerin analiz edilmesi radyologlar için ciddi bir yüktür. Bir BT taramasındaki görüntü sayısı 600'e kadar çıkabilmektedir. Bu sebeple bilgisayar destekli tespit sistemleri görüntülerin daha hızlı ve daha doğru tanınması için çok önemlidir. Bu çalışmada evrişimsel sinir ağları (ESA) kullanılarak akciğer BT görüntülerini analiz eden ve yanlış-pozitifleri azaltan bir yöntem geliştirilmiştir. Sinir ağı modeli, farklı boyutlardaki girdiler ile denenmiş ve girdi boyutunun sistem performansına olan etkisi gösterilmiştir. Ayrıca, bir çok modelden elde edilen sonuçlar değişik kombinasyonlarda bir araya getirilerek başarım artırılmış ve bu yöntemin gücü gösterilmiştir. Sınıflandırılacak bilginin 3 boyutlu olması ve veriyi 2 boyutlu işlemenin bilgi kaybına yol açmasından dolayı 3 boyutlu evrişimsel sinir ağları kullanılmıştır. Önerilen yöntem LUNA16 Yarışması tarafından sağlanan veri seti üzerinde denenmiş ve tarama başına 1 yanlış pozitif oranında 0.831 duyarlılığına ulaşılmıştır.

*Anahtar Kelimeler — Akciğer Nodül Tespiti, Bilgisayarlı Tomografi, Evrişimsel Sinir Ağları, Derin Öğrenme.*

*Abstract—* Recent studies have shown that lung cancer screening using annual low-dose computed tomography (CT) reduces lung cancer mortality by 20% compared to traditional chest radiography. Therefore, CT lung screening has started to be used widely all across the world. However, analyzing these images is a serious burden for radiologists. The number of slices in a CT scan can be up to 600. Therefore, computer-aided-detection (CAD) systems are very important for faster and more accurate assessment of the data. In this study, we proposed a framework that analyzes CT lung screenings using convolutional neural networks (CNNs) to reduce false positives. We trained our model with different volume sizes and showed that volume size plays a critical role in the performance of the system. We also used different fusions in order to show their power and effect on the overall accuracy. 3D CNNs were preferred over 2D CNNs because 2D convolutional operations applied to 3D data could result in information loss. The proposed framework has been tested on the dataset provided by the LUNA16 Challenge and resulted in a sensitivity of 0.831 at 1 false positive per scan.

## I. GİRİŞ

Akciğer kanseri, akciğer dokularındaki kötü huylu hücrelerin kontrolsüz büyümesi ile oluşmaktadır. Bu kanser türü 2012 yılında dünya çapında 1.8 milyon kişide görülmüş ve bu vakaların 1.6 milyonu ölüm ile sonuçlanmıştır [1]. Bu istatistiklerle, kansere dayalı ölümlerde erkeklerde ilk sırada, kadınlarda ise meme kanserinden sonra ikinci sıradadır [1]. ABD'de gerçekleştirilen ve 50.000'den fazla yüksek riskli denek içeren Ulusal Akciğer Tarama Testi (NLST), düşük doz BT'nin akciğer radyolojisine göre erken teşhiste %20 daha avantajlı olduğunu ortaya çıkartmıştır [2]. Bu sebeple akciğer BT taramaları tüm dünyada yaygınlaşmaktadır.

BT'nin en büyük zorluklarından biri, çok fazla sayıda görüntünün radyologlar tarafından detaylı incelenmesini gerektirmesidir. BT taramasındaki görüntü sayısı 600'e kadar çıkabilmektedir. Bu büyüklükteki verilerin analiz edilmesi radyologlar için ciddi bir yük oluşturmaktadır. Bu sebeple, bilgisayar destekli tespit (BDT) sistemleri verilerin daha hızlı ve doğru tespit edilmesinde çok önemlidir. Geçtiğimiz yirmi yılda bu konuyla ilgili birçok çalışma gerçekleştirilmiştir. Yapılan bu çalışmalara bakıldığında BDT sistemlerinin iki adım içerdiği gözlemektedir: 1) nodül adaylarının tespiti, 2) yanlış pozitiflerin azaltılması. Nodül adayı üretiminin amacı nodül olma potansiyeli taşıyabilecek her türlü konumu tespit etmektir. Bu adımdaki yüksek duyarlılık çok önemlidir. Amaç hiçbir nodülü kaçırmamaktır bu yüzden nodül olmayan bir çok nokta da algoritmalar tarafından üretilmektedir. Yanlış pozitiflerin azaltılması aşamasının amacı aynı hassasiyet korunarak ilk adımda üretilmiş olan yanlış nodül işaretlemelerinin azaltılmasını sağlamaktır.

Evrişimsel sinir ağları (ESA), birçok alandaki yüksek performansı sebebiyle son yıllarda oldukça popüler hale gelmiştir. ESA, ağırlıklarının sistem tarafından öğrenildiği birçok sinir düğümünden oluşmaktadır. ESA'nın geleneksel yapay sinir ağlarına göre en önemli avantajlarından biri filtreleme (evrişim) operasyonları ile özniteliklerin kendi kendine çıkarılmasıdır. Bu filtreleme operasyonları veri üzerindeki uzamsal ilişkiyi çıkarıp modellemeye dâhil etmektedir, bu sebeple nesne tanıma, video analizi, doğal dil işleme ve ses tanıma gibi birçok görevde ESA çok yüksek performans yakalamaktadır [3]. Geçtiğimiz yıllarda biyomedikal görüntüler üzerinde 3B ESA yöntemi kullanılmış olup başarılı sonuçlar elde edilmiştir [4] [5] [6] [7].

Bu çalışmadaki amaç nodül adaylarını tespit eden algoritmalar tarafından üretilmiş noktalardaki yanlış pozitifleri azaltmak ve ESA modelinde girdi veri boyutu değişiminin sonuca olan etkisini göstermektir. Kullanılan veri tabanı LUNA16: Lung Nodule Analysis Challenge [8] tarafından sağlanmıştır.

Veri setindeki nodül boyutları büyük farklılık göstermektedir. En küçük nodül 3 mm çapında, en büyük nodül ise 34 mm çapındadır. Bu durum ESA model eğitiminde bir problem oluşturmaktadır çünkü modelin girdi boyutu bütün örnekler için aynı olmalıdır. Eğer küçük bir girdi boyutu seçilirse model büyük nodülleri görmemiş ve dolayısıyla onları öğrenememiş olacaktır. Büyük bir girdi boyutu seçildiği durumda ise küçük nodüllerin etrafındaki parçacıklar gürültü yaratacaklardır; bu da modelin küçük nodülleri öğrenmesini zorlaştıracaktır. Literatür incelendiğinde farklı girdi boyutlarını benzer modeller üzerinde inceleyen bir makale bulunmamaktadır. Bu makaledeki en önemli katkı farklı girdi boyutlarının birbirine çok benzer ESA modelleri üzerindeki performansını karşılaştırmak ve karar füzyonu (decision fusion) kullanarak performansın artışını göstermektir.

## II. METOT

Bu çalışmada önerilen yöntemde, farklı çözünürlükteki taramalardan kaynaklanacak problemleri yok etmek için öncelikle tüm veriler yeniden ölçeklendirilmiştir. Sınıf sayılarındaki orantısızlığı yok etmek için farklı bir minik-toptan (mini-batch) yapısı kullanılmıştır. Modelin genelleştirme kabiliyetini artırmak için dışarıda-tut (dropout) yöntemi uygulanmıştır. Farklı girdi boyutları, birbirine çok benzer modellerde denenmiş, karar füzyonu ile performans artırılmıştır.

### A. Veri Seti

Sağlanan veri setinde 888 adet BT taraması bulunmaktadır. Her BT taramasındaki görüntü dilimi sayısı 120 ile 600 arasında değişmektedir. Veri setinde toplamda 1188 adet nodül bulunmaktadır. Fakat sağlanan etiket listesinde bu nodüllerin sadece 1166 tanesi bulunmaktadır. Toplamda 754,975 nodül adayı işaretlenmiştir. Bu aday noktalar birçok aday nokta üretme algoritmasının sonuçlarının birleştirilmesiyle oluşturulmuştur. 1166 nodüle karşılık 1557 nokta nodül olarak işaretlenmiştir (bazı nodüller farklı kesitlerden saptanmıştır). Geri kalan 753,418 nokta ise yanlış pozitiftir.

### B. Ön İşleme

Farklı tıbbi cihazlar ve tarama özelliğinden ötürü veri setindeki BT taramalarının çözünürlüğü birbirinden farklıdır. Bu durum ESA modeli için sorun yaratmaktadır çünkü belirli bir girdi boyutu gerçek dünyada farklı boyutlara tekabül edecektir. Örneğin 18x20x20 piksel boyutlarındaki bir matris, bir BT taraması için 30x40x40 milimetreye tekabül edebilecekken başka bir BT taraması için 24x36x36 boyutlarına tekabül edebilir. ESA modeli için girdilerin homojen bir yapıda olması performansı artıracaktır. Bu sebeple bütün BT taramalarındaki çözünürlük aynı olacak şekilde bütün veriler yeniden ölçeklendirilmiştir. Doğru çözünürlüğü bulmak için veri setindeki bütün taramalardaki pikseller arasındaki mesafeler incelenmiştir. X, Y ve Z eksenlerindeki pikseller arası mesafelerin histogramı Resim 1'de verilmiştir.

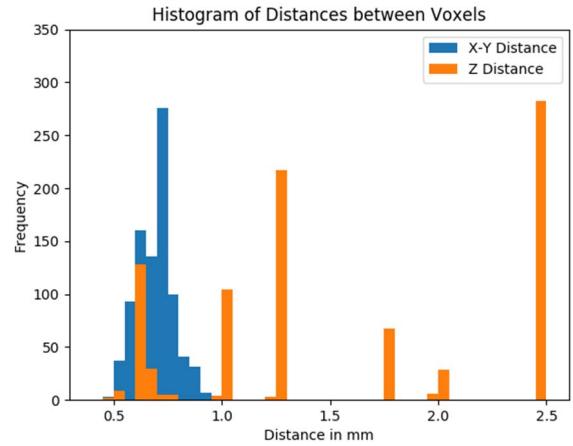

**Resim 1:** X, Y ve Z eksenlerinde iki piksel arasındaki uzaklıkların dağılımı.

Bu histograma göre, X ve Y eksenleri için iki piksel arasındaki ortalama mesafe 0.69 milimetreyken, Z ekseni için bu mesafe 1.56 milimetredir. Bu bilgi doğrultusunda bütün taramalar pikseller arasındaki mesafe 0.7x0.7x1 (X, Y ve Z) milimetre olacak şekilde yeniden ölçeklendirilmiştir. Z ekseni ortalamadan daha kısa bir mesafeye sabitlenerek bu eksende çözünürlüğün artırılması hedeflenmiştir.

### C. Eğitim

Saklı katmandaki düğüm sayılarındaki farklılık dışında birbirinin aynısı 5 model belirlenmiştir ve sonuçlar karşılaştırılmıştır. Farklı girdi boyutları veri setindeki nodüllerin farklı karakteristiklerine odaklanmaktadır. Örneğin girdi boyutu küçük ise, model küçük nodülleri büyüklere göre daha iyi öğrenmektedir. Diğer yandan eğer girdi boyutu büyük ise model bütün nodülleri kapsamaktadır fakat küçük nodüllerin öğrenilmesi daha da zorlaşmaktadır. Nodüllerin büyüklüklerinin dağılımı Doi tarafından yayımlanan çalışmada verilmiştir [9]. Bu dağılıma göre, 12x24x24 piksel boyutu bütün nodüllerin yaklaşık %80'ini kapsamaktadır; 36x48x48 piksel boyutu ise bütün nodülleri kapsayan en küçük boyuttur. Bu dağılım göz önüne alındığında bu çalışmada kullanılmak üzere 12x24x24, 18x30x30, 24x36x36, 30x42x42 ve 36x48x48 piksel boyutları belirlenmiştir. Girdi boyutlarının farklı olması

**Tablo 1:** ESA modellerinin mimarileri.

| Model-1 | Model-2 | Model-3 | Model-4 | Model-5 |
|---|---|---|---|---|
| Girdi Boyutu: 12x24x24 | Girdi Boyutu: 18x30x30 | Girdi Boyutu: 24x36x36 | Girdi Boyutu: 30x42x42 | Girdi Boyutu: 36x48x48 |
| C1 64@3x5x5 | C1 64@3x5x5 | C1 64@3x5x5 | C1 64@3x5x5 | C1 64@3x5x5 |
| MaxPool (3,3,3) | MaxPool (3,3,3) | MaxPool (3,3,3) | MaxPool (3,3,3) | MaxPool (3,3,3) |
| Dropout 0.2 | Dropout 0.2 | Dropout 0.2 | Dropout 0.2 | Dropout 0.2 |
| C2 64@3x5x5 | C2 64@3x5x5 | C2 64@3x5x5 | C2 64@3x5x5 | C2 64@3x5x5 |
| MaxPool (2, 2, 2) | MaxPool (2, 2, 2) | MaxPool (2, 2, 2) | MaxPool (2, 2, 2) | MaxPool (2, 2, 2) |
| Dropout 0.2 | Dropout 0.2 | Dropout 0.2 | Dropout 0.2 | Dropout 0.2 |
| Saklı Katman: 150 düğüm | Saklı Katman: 250 düğüm | Saklı Katman: 350 düğüm | Saklı Katman: 400 düğüm | Saklı Katman: 600 düğüm |
| Çıktı Katmanı: 2 düğüm | Çıktı Katmanı: 2 düğüm | Çıktı Katmanı: 2 düğüm | Çıktı Katmanı: 2 düğüm | Çıktı Katmanı: 2 düğüm |
| SoftMax | SoftMax | SoftMax | SoftMax | SoftMax |

tam bağlı katmandaki düğüm sayısını doğrudan etkilemektedir. Saklı katmandaki düğüm sayısı, doğrudan sistemdeki düğüm sayısı ile orantılı olduğu için önerilen modellerde bu kısım her model için ayrıca belirlenmiştir. Bu çalışmada karşılaştırılan modeller ve mimarileri Tablo 1'de verilmiştir. Nodül örneklerini çoğaltmak için makine öğrenmesi alanında sıkça kullanılan veri çoğaltma (data augmentation) yöntemi kullanılmıştır. Her nodülün 3 eksene göre simetrisi alınmış, aynı zamanda görüntüler 90, 180 ve 270 derecelerde döndürülmüştür. Bu yöntem ile nodüllerin sayısı 16 kat artmıştır.

Karar füzyonu makine öğrenimi problemlerinde performansı artırmak için uygulanan yöntemlerden biridir. Yaygın olarak kullanılan metotlardan biri, her bir örnek veri için her modelin ürettiği olasılık tahminlerinin ortalamasını almaktır. Bu çalışmada, 5 farklı modelin sonuçları çeşitli birleşimlerde bir araya getirilmiş ve sonuçlar gözlemlenmiştir.

Veri setinde doğru pozitiflerin yanlış pozitiflere oranı yaklaşık 1'e 483'tür. Sınıf dağılımlarının bu şekilde orantısız olduğu veri setlerinde model az sayıdaki sınıfı öğrenememektedir çünkü eğitim sırasında diğer sınıfa oranla çok daha az karşılaşmış olunmaktadır. Bu durumu çözmek için şu şekilde bir algoritma geliştirilmiştir:

1. Eğitim setindeki tüm doğru pozitifleri topla. Bunların sayısı N olsun.
2. N adet yanlış pozitif topla.
3. Bu iki seti ortak bir kümede topla, karıştır ve bu kümeyi eğit.
4. Yeni bir N adet yanlış pozitif topla, doğru pozitifler ile karıştır ve yeniden eğit.
5. Eğitim işlemi bitene kadar 4. adımı uygula.

Bu algoritma ile eğitim sırasında iki sınıf tarafındaki örnekler ile eşit sayıda karşılaşma olmaktadır. Eğitim yapılırken minitoptan sayısı 32 olarak belirlenmiştir. Optimizasyon algoritması olarak Adam kullanılmıştır [10]. Modelin uygulanması Keras [11] derin öğrenme kütüphanesi kullanılarak Python programlama dili ile gerçekleştirilmiştir. Algoritma NVIDIA TITAN X GPU üzerinde koşturulmuştur. Veri seti üzerinde 10 kat çapraz geçerleme (cross-validation) uygulanmıştır.

*D. Değerlendirme*

Yarışma kapsamında değerlendirme, FROC (Free-Receiver Operating Characteristics) eğrisi üzerindeki 7 farklı yanlış pozitif oranındaki (1/8, 1/4, 1/2, 1, 2, 4 ve 8) duyarlılıkların ortalamasının alınması ile oluşturulmaktadır. FROC eğrisi bu tarz problemlerde eşik değerini belirlemek için kullanılan ve sonucun farklı eşik değerlerindeki performansını gösteren bir grafiktir.

## III. DENEY SONUÇLARI

Çalışmaya dahil edilen 5 farklı model için deney sonuçları Tablo 2'deki gibidir. Sonuçlar göstermektedir ki, birbirinin neredeyse aynısı modellerde bile farklı girdi boyutları performansta önemli bir fark yaratmaktadır. Örneğin, Model-1 ve Model-5 arasında 0.107 puan fark bulunmaktadır.

İkinci olarak, farklı model birleşimleri karar füzyonu için kullanılmıştır. Bu birleşimler sırasıyla şu şekildedir:

Füzyon-1: Model-1, Model-3, Model-5
Füzyon-2: Model-1, Model-2, Model-5
Füzyon-3: Model-1, Model-4, Model-5
Füzyon-4: Model-1, Model-2, Model-3, Model-4, Model-5

**Tablo 2:** 5 farklı modelin 7 yanlış pozitif oranındaki duyarlılıkları.

| Yanlış Pozitif Oranı | DUYARLILIK | | | | |
|---|---|---|---|---|---|
| | M1 | M2 | M3 | M4 | M5 |
| **0.125** | 0.431 | 0.444 | 0.51 | 0.547 | 0.591 |
| **0.25** | 0.504 | 0.537 | 0.613 | 0.632 | 0.676 |
| **0.5** | 0.593 | 0.631 | 0.689 | 0.708 | 0.736 |
| **1** | 0.68 | 0.709 | 0.76 | 0.78 | 0.779 |
| **2** | 0.739 | 0.763 | 0.815 | 0.825 | 0.822 |
| **4** | 0.79 | 0.811 | 0.868 | 0.858 | 0.847 |
| **8** | 0.844 | 0.856 | 0.9 | 0.89 | 0.879 |
| **Ortalama** | 0.654 | 0.679 | 0.736 | 0.749 | 0.761 |

Tablo-3 füzyonların performansı nasıl artırdığını göstermektedir. En iyi sonuç bütün modellerin birleşimi ile oluşmaktadır.

**Tablo 3:** 5 modelin 4 farklı füzyonunun 7 yanlış pozitif oranındaki duyarlılıkları.

| Yanlış Pozitif Oranı | DUYARLILIK | | | |
|---|---|---|---|---|
| | F1 | F2 | F3 | F4 |
| **0.125** | 0.57 | 0.548 | 0.577 | 0.588 |
| **0.25** | 0.654 | 0.633 | 0.681 | 0.669 |
| **0.5** | 0.737 | 0.722 | 0.761 | 0.749 |
| **1** | 0.802 | 0.784 | 0.815 | 0.831 |
| **2** | 0.855 | 0.84 | 0.86 | 0.863 |
| **4** | 0.89 | 0.873 | 0.887 | 0.892 |
| **8** | 0.915 | 0.892 | 0.904 | 0.913 |
| **Ortalama** | 0.775 | 0.756 | 0.784 | 0.786 |

Resim 2'de 5 farklı modelin yanlış pozitif oranları ve oranlara karşılık gelen duyarlılıklar FROC eğrisi üzerinde gösterilmiştir. Sınıflandırma için eşik eğrisi düştüğünde duyarlılığın arttığı, ancak aynı zamanda yanlış pozitiflerin de arttığı gözlemlenmektedir.

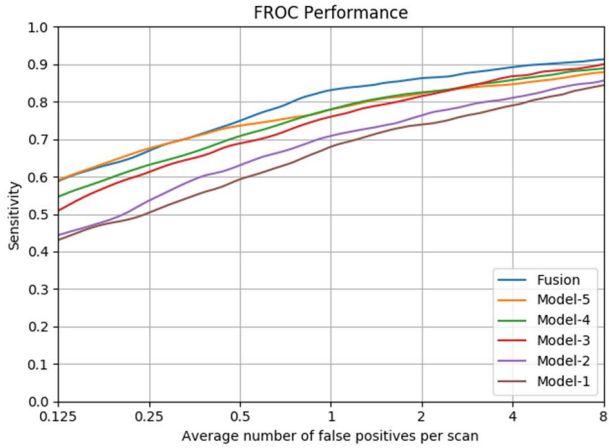

**Resim 2:** 5 farklı model ve Füzyon-4 için FROC eğrileri.

## IV. LUNA16 YARIŞMASI

Birçok firma ve üniversitelerdeki araştırma grubu LUNA16 yarışmasına katılmıştır. LUNA16 yarışması tarafından sağlanan veri seti bu alanda sağlanan en büyük veri setidir bu yüzden farklı yöntemlerin aynı set üzerinde karşılaştırılması için oldukça uygundur. Bu yarışma ile ilgili sıkıntılı nokta test setinin de açık olmasıdır. Değerlendirmelerde, bütün veri seti 10 kümeye bölünüp, her seferinde 9 set eğitim seti, geri kalan set ise test kümesi olarak kullanılmaktadır. Son zamanlarda yüksek performanslı başvurularının olması sebebiyle organizatörler bir açıklama yapmış ve bazı başvuruların "Gerçek olamayacak kadar yüksek performanslı" olduğunu belirtmiştir. Aynı zamanda bazı başvuruların da veri setine göre ayarlandığını bu sebeple gerçeği yansıtmadığını söylemişlerdir [11]. Bu çalışmada elde edilen sonuçtan daha iyi performansa sahip iki akademik çalışma vardır [9] [13]. Bu çalışmalar incelendiğinde daha fazla katman kullanıldığı ve farklı veri çoğaltma uygulandığı gözlemlenmektedir.

## V. YORUM

Elde edilen sonuçlar, iyi bir ön işleme ve değerleri optimize edilmiş bir ESA modeli ile 3 boyutlu veri üzerinde nodül sınıflandırılmasında iyi sonuç elde edilebildiğini göstermiştir. Aynı zamanda hedef nodüllerin boyutlarının oldukça değişken olduğu bu tarz problemlerde, farklı girdi boyutlarının avantajı kullanılarak performans daha da artırılabilmektedir. Farklı girdi boyutları eğitim sırasında farklı noktalara odaklanılmasını sağlamaktadır, bu sebeple farklı güçlü yanları oluşmaktadır. Karar füzyonu ile farklı modellerin güçlü yanları bir araya getirilip sonuç daha da ileriye taşınabilmektedir. Karar füzyonu ile sonuç en iyi modele göre 0.25 puan artırılmıştır.

İlerideki çalışmalarda daha derin ESA yapıları kullanılarak performansın daha da artırılması mümkündür. Daha çok evrişimsel katmanın olduğu mimariler daha karışık modelleri öğrenebilmektedir.

## KAYNAKLAR

[1] World Health Organization, World Cancer Report 2014, Lyon: International Agency For Research on Cancer, 2014.

[2] D. R. Arbele, A. M. Adams, C. D. Berg, W. C. Black, J. D. Clapp, R. M. FagerStrom, I. F. Gareen, C. Gatsonis, P. M. Marcus ve J. D. Sicks, «Reduced lung-cancer mortality with low-dose computed tomographic screening.,» *N Engl J Med*, cilt 365, no. 5, pp. 395-409, 2011.

[3] A. Bhandare, M. Bhide, P. Gokhale ve R. Chandavarkar, «Applications of Convolutional Neural Networks,» *International Journal of Computer Science and Information Technologies*, cilt 7, no. 5, pp. 2206-2215, 2016.

[4] Çiçek et al., «3D U-Net: Learning Dense Volumetric Segmentation from Sparse Annotation,» [Çevrimiçi]. Available: https://arxiv.org/abs/1606.06650.

[5] Dou et al., «Automatic detection of cerebral microbleeds from MR images via 3D convolutional nerual networks,» *IEEE Transactions on medical Imaging*, cilt 35, no. 5, pp. 1182-1195, 2016.

[6] Dou et al., «3D deeply supervised network for automatic liver segmentation from CT volumes.,» 2016. [Çevrimiçi]. Available: https://arxiv.org/abs/1607.00582.

[7] Chen et al., «Voxresnet: Deep voxelwise residual networks for volumetric brain segmentation,» 2016. [Çevrimiçi]. Available: https://arxiv.org/abs/1608.05895.

[8] «LUNA16 Challenge,» [Çevrimiçi]. Available: https://luna16.grand-challenge.org/. [Erişildi: October 2017].

[9] D. Qi, C. Hao, Y. Lequan, Q. Jing ve H. Pheng-Ann, «Multilevel Contextual 3-D CNNs for False Positive Reduction in Pulmonary Nodule Detection,» *IEEE Transactions on biomedical Engineering*, cilt 64, no. 7, pp. 1558-1567, 2017.

[10] D. P. Kingma ve J. Ba, «Adam: A Method for Stochastic Optimization,» %1 içinde *3rd International Conference for Learning Representations*, San Diego, 2015.

[11] «Keras Documentation,» [Çevrimiçi]. Available: https://keras.io/.

[12] «LUNA16: Lung Nodule Analysis,» 2017. [Çevrimiçi]. Available: https://luna16.grand-challenge.org/announcements/. [Erişildi: December 2017].

[13] A. A. A. Setio, F. Ciompi, G. Litjens, P. Gerke, C. Jacobs, S. J. van Riel ve M. M. W. Wille, «Pulmonary Nodule Detection in CT Images: False Positive Reduction Using Multi-View Convolutional Networks,» *IEEE Transactions on Medical Imaging*, cilt 35, no. 5, pp. 1160-1169, 2016.